# Universal approximation using differentiators and application to feedback control

Xinhua Wang

Advanced Control Technology, Department of Electrical & Computer Engineering, National University of Singapore, Singapore 117576

E-mail: wangxinhua@buaa.edu.cn

**Abstract:** In this paper, we consider the problems of approximating uncertainties and feedback control for a class of nonlinear systems without full-known states, and two approximation methods are proposed: universal approximation using integral-chain differentiator or extended observer. Comparing to the approximations by fuzzy system and radial-based-function (RBF) neural networks, the presented two methods can not only approximate universally the uncertainties, but also estimate the unknown states. Moreover, the integral-chain differentiator can restrain noises thoroughly. The theoretical results are confirmed by computer simulations for feedback control.

**Keywords:** Universal approximation, integral-chain differentiator, uncertainties.

## 1. Introduction

This paper focuses on the problem of universally approximating uncertainties for uncertain nonlinear systems without full-known states. Usually the uncertainties of nonlinear systems are approximated by some intelligent algorithms, such as fuzzy systems or radial-based-function (RBF) networks. Over the last decade, control researchers tried to apply intelligent methodologies to the design of estimation and control for uncertain nonlinear systems [1]-[11]. In paper [1], a conclusion was given that fuzzy systems are universal approximation. The Stone-Weierstrass Theorem is used to prove that fuzzy systems with product inference, centroid defuzzification, and Gaussian membership function are capable of approximating any real continuous function on a compact set to arbitrary accuracy. Paper [2] reported on a related study of radial-basis-function (RBF) networks, and it was proved that RBF networks having one hidden layer are capable of universal approximation. The main difficulties with estimating uncertainties by these algorithms are: 1) the parameters or the neural network weights are difficult to be regulated; 2) membership function and Gaussian function are selected by experiences; 3) all of the system states must be required for estimation and feedback; 4) noises cannot be restrained by the above approximation algorithms.

The development of differentiator provides an effective method for signal tracking [12-18]. Using differentiator, we can estimate the derivatives of a signal. One of the important merits of the estimation of the derivatives of signal by differentiator is not based on the model of the system. We know that most systems are described by differential equations and the uncertainties of system exist in the differential equations. Usually, these differential equations can be transferred to the integral-chain forms, and the differentiator can estimate the derivatives of the states. By obtaining the derivatives of states, the differential equation becomes an algebraic equation. Therefore, the uncertainties can be carried out in algebraic equation directly. Moreover, the uncertainties can be taken as new unknown states of the system, and an extended observer can be designed to estimate all the unknown states including the uncertainties.

Differentiation of signals is an old and well-known problem [12]–[14] and has attracted more attention in recent years [15]–[18]. In [15, 16], a differentiator via second-order (or high-order) sliding modes algorithm has been proposed. The information one needs to know on the signal is the upper bounds for Lipschitz constant of the derivatives of the signal. It constrains the types of input signals. And for this differentiator, the chattering phenomenon is inevitable. The popular high-gain differentiators in [17, 18] provide for an exact derivative when their gains tend to infinity. Noises exist in each layer of differential equations, therefore, all the derivatives estimations of input signal are all affected by noises directly.

In [19], we presented a finite-time-convergent differentiator based on singular perturbation technique. However, the differentiators are complicated and difficult to be implemented in practice, due to the long computation time. In [20], we designed a nonlinear tracking-differentiator with high speed, and it succeeds in application to velocity estimation for low-speed regions only based on position measurements [21]. However, it only can obtain first-order derivative of a signal not arbitrary-order derivatives.

This paper provides an output feedback tracking control method for uncertain system without full-known states. The new result is facilitated by a high-order integral-chain differentiator. The proposed differentiator algorithm has an integral-chain structure and can not only approximate uncertainties, but also estimate the unknown states. It is shown that, integral-chain differentiator can restrain noises more thoroughly than usual high-gain linear differentiator. In integral-chain differentiator, disturbances only exist in the last differential equation and can be restrained through each layer of integrator. An extended observer can be designed to estimate all the unknown states including the uncertainties under the condition that the uncertainties are taken as new unknown states of the system. Moreover, an output feedback controller based on integral-chain differentiator is designed to stabilize the uncertain nonlinear system without full-known states.

This paper is organized in the following format. In Section 2, problem statement is given, and the approximations by fuzzy systems and RBF networks are recalled. In Section 3, output feedback control based on universal approximations using integral-chain differentiator is presented. In Section 4, output feedback control based on extended observer is designed. In Section 5, the simulations are given, and our conclusions are made in Section 6.

## 2. Problem statement

In the following, we will recall the approximation algorithms by fuzzy systems and RBF networks respectively [1-11], and a full-state feedback control for the nonlinear system is introduced firstly.

Consider $n$-order nonlinear system as follow:

$$\dot{x}_1 = x_2$$
$$\cdots$$
$$\dot{x}_{n-1} = x_n \qquad (1)$$
$$\dot{x}_n = f(x) + g(x)u$$
$$y = x_1$$

where $x=(x_1,\ldots, x_n)^\mathrm{T}$ is the state vector, $f$ and $g$ are all nonlinear functions, $u \in R$ and $y \in R$ are the input control and output respectively. Function $g(x)$ is bounded, i.e.,

$$l_{\inf} \leq |g(x)| \leq l_{\sup}$$

where $l_{\inf}$ and $l_{\sup}$ are positive constants. Let the desired output be $y_d$, and denote

$$e_i = x_i - y_d^{(i-1)}, \quad i=1,\cdots,n, \quad e=(e_1,\cdots,e_n)^\mathrm{T} \qquad (2)$$

Select $K=(k_1,\cdots,k_n)$ such that $s^n + k_n s^{(n-1)} + \cdots + k_1 = 0$ is Hurwitz. Therefore the tracking error system is

$$\dot{e}_1 = e_2$$
$$\cdots$$
$$\dot{e}_{n-1} = e_n$$
$$\dot{e}_n = f(x) + g(x)u - y_d^{(n)}$$

If the nonlinear function $f(x)$ and the state vector $x$ are all known, we can select the control law as

$$u = \frac{1}{g(x)}\left[-f(x) + y_d^{(n)} + Ke\right] \qquad (3)$$

From (1) and (3), we have

$$\dot{e}_n + k_n e_n + \cdots + k_1 e_1 = 0 \tag{4}$$

Therefore,

$$e_i(t) \to 0, \quad i = 1, \cdots, n$$

for $t \to \infty$.

If $f(x)$ and $x_2, \ldots, x_n$ are unknown, we should estimate $f(x)$ by some algorithms. In the following, we will simply recall fuzzy systems and neural networks respectively to approximate uncertain item $f(x)$.

**2.1. Controller design based on fuzzy system**

**2.1.1. Uncertainty approximation using fuzzy system**

If $f(x)$ is unknown, we can replace $f(x)$ with the fuzzy estimation $\hat{f}(x)$ to realize feedback control.

*Step one:* For $x_i$ ($i=1,2,\cdots,n$), define the fuzzy sets $A_i^{l_i}$, $l_i = 1, 2, \cdots, p_i$.

*Step two:* Adopt $\prod_{i=1}^{n} p_i$ fuzzy rules to construct fuzzy system $\hat{f}(x|\theta_f)$:

$$R^{(j)}: \text{If } x_1 \text{ is } A_1^{l_1} \text{ and } \ldots \text{ and } x_n \text{ is } A_1^{l_n} \text{ then } \hat{f} \text{ is } E^{l_1 \cdots l_n} \tag{5}$$

where $l_i = 1, 2, \cdots, p_i$, $i = 1, 2, \cdots, n$. Therefore, the output of fuzzy system is

$$\hat{f}(x|\theta_f) = \frac{\sum_{l_1=1}^{p_1} \cdots \sum_{l_n=1}^{p_n} \bar{y}_f^{l_1 \cdots l_n} \left( \prod_{i=1}^{n} \mu_{A_i^{l_i}}(x_i) \right)}{\sum_{l_1=1}^{p_1} \cdots \sum_{l_n=1}^{p_n} \left( \prod_{i=1}^{n} \mu_{A_i^{l_i}}(x_i) \right)} \tag{6}$$

where $\mu_{A_i^j}(x_i)$ is the membership function of $x_i$. All the states are required known. Moreover, if there are noises in the measurement output $y = x_1$, the computation of $\mu_{A_i^j}(x_i)$ is affected seriously, therefore, fuzzy system is contaminated.

Let $\bar{y}_f^{l_1 \cdots l_n}$ be a free parameter and be put in the set $\theta_f \in R^{\prod_{i=1}^{n} p_i}$. Column vector $\xi(x)$ is introduced and (6) can be written as:

$$\hat{f}(x|\theta_f) = \theta_f^T \xi(x) \tag{7}$$

where $\xi(x)$ is the $\prod_{i=1}^{n} p_i$-dimensional column vector, and $l_1 \cdots, l_n$ elements are respectively

$$\xi_{l_1 \cdots l_n}(x) = \frac{\prod_{i=1}^{n} \mu_{A_i^{l_i}}(x_i)}{\sum_{l_1=1}^{p_1} \cdots \sum_{l_n=1}^{p_n} \left( \prod_{i=1}^{n} \mu_{A_i^{l_i}}(x_i) \right)} \tag{8}$$

The membership functions are needed to be selected according to experiences. Moreover, all the states must be known,

and $\hat{f}(x|\theta_f)$ is affected easily by noises.

### 2.1.2. Design of adaptive fuzzy controller

The fuzzy system is used to approximate $f$, and the control input (3) is written as

$$u = \frac{1}{g(x)}\left[-\hat{f}(x|\theta_f) + y_d^{(n)} + K^T e\right] \tag{9}$$

$$\hat{f}(x|\theta_f) = \theta_f^T \xi(x) \tag{10}$$

where $\xi(x)$ is the fuzzy vector, and the adaptive rule is selected as

$$\dot{\theta}_f = -\gamma_1 e^T P b \xi(x) \tag{11}$$

### 2.1.3. Stability analysis of adaptive fuzzy controller

From (1) and (9), we get the close-loop error system as follow:

$$e^{(n)} = -K^T e + \left[\hat{f}(x|\theta_f) - f(x)\right] \tag{12}$$

Let

$$\Lambda = \begin{bmatrix} 0 & 1 & 0 & 0 & \cdots & 0 & 0 \\ 0 & 0 & 1 & 0 & \cdots & 0 & 0 \\ \cdots & \cdots & \cdots & \cdots & \cdots & \cdots & \cdots \\ 0 & 0 & 0 & 0 & \cdots & 0 & 1 \\ -k_n & -k_{n-1} & \cdots & \cdots & \cdots & \cdots & -k_1 \end{bmatrix}, \quad b = \begin{bmatrix} 0 \\ 0 \\ \cdots \\ 0 \\ 1 \end{bmatrix} \tag{13}$$

Therefore, we rewrite (12) as

$$\dot{e} = \Lambda e + b\left[\hat{f}(x|\theta_f) - f(x)\right] \tag{14}$$

Suppose the optimal parameter as

$$\theta_f^* = \arg \min_{\theta_f \in \Omega_f} \left[\sup_{x \in R^n} \left|\hat{f}(x|\theta_f) - f(x)\right|\right] \tag{15}$$

where $\Omega_f$ is the set of $\theta_f$, i.e., $\theta_f \in \Omega_f$.

Define minimum approximation error as

$$\omega = \hat{f}(x|\theta_f^*) - f(x) \tag{16}$$

Therefore, (14) can be written as

$$\dot{e} = \Lambda e + b\left\{\left[\hat{f}(x|\theta_f) - \hat{f}(x|\theta_f^*)\right] + \omega\right\} \tag{17}$$

From (10) and (17), we have

$$\dot{e} = \Lambda e + b\left[(\theta_f - \theta_f^*)^T \xi(x) + \omega\right] \tag{18}$$

Equation (18) describes the relation of tracking error and $\theta_f$. The task of adaptive rule is to determine a regulation mechanism for $\theta_f$ and make the tracking error $e$ and $\theta_f - \theta_f^*$ minimum.

Define the Lyapunov function as

$$V = \frac{1}{2}e^T Pe + \frac{1}{2\gamma_1}\left(\theta_f - \theta_f^*\right)^T \left(\theta_f - \theta_f^*\right) \tag{19}$$

where $\gamma_1$ is a positive constant, $P$ is a positive-defined matrix and satisfied the following Lyapunov function

$$\Lambda^T P + P\Lambda = -Q \tag{20}$$

where $Q$ is an arbitrary $n \times n$ positive-defined matrix, and $\Lambda$ is defined in (13).

The derivative of $V$ is

$$\dot{V} = -\frac{1}{2}e^T Qe + e^T Pb\omega + \frac{1}{\gamma_1}\left(\theta_f - \theta_f^*\right)^T \left[\dot{\theta}_f + \gamma_1 e^T Pb\xi(x)\right] \tag{21}$$

From (11), we have

$$\dot{V} = -\frac{1}{2}e^T Qe + e^T Pb\omega \tag{22}$$

Due to the approximation error $\omega$ is sufficiently small, we can obtain approximately $\dot{V} \leq 0$.

## 2.2. Adaptive control based on RBF networks
### 2.2.1. Basic neural network

RBF networks are used to approximate adaptively the uncertain $f$. The algorithm of RBF networks are:

$$h_j = g\left(\|x - c_{ij}\|^2 / b_j^2\right)$$

$$f = W^T h(x) + \varepsilon$$

where $x$ is the input signal of the network, $i$ is the input number of the network, $j$ is the number of hidden layer nodes in the network, $h = [h_1, h_2, \cdots, h_n]^T$ is the output of Gaussian function, $W$ is the neural network weights, $\varepsilon$ is approximation error of neural network, and $\varepsilon \leq \varepsilon_N$.

RBF network approximation $f$ is used. The network input is selected as $x = [e \ \dot{e}]^T$, and the output of RBF neural network is

$$\hat{f}(x) = \hat{W}^T h(x) \tag{23}$$

where $h(x)$ is the Gaussian function of neural network.

We know that Gaussian function and the neural network weights are difficult to be selected.

### 2.2.2. Design and analysis of adaptive neural network controller

Neural network approximation (23) is used, and the control rule (3) can be written as (9) with (23).
Let adaptive rule be

$$\dot{\hat{W}} = -\gamma e^T Pbh(x) \tag{24}$$

From (1) and (9), we have

$$e^{(n)} = -K^T e + \left[\widehat{f}(x) - f(x)\right] \tag{25}$$

$\Lambda$ and $b$ are defined in (13). Therefore, (25) can be written as

$$\dot{e} = \Lambda e + b\left[\widehat{f}(x) - f(x)\right] \tag{26}$$

Let optimal parameter be

$$W^* = \arg\min_{W \in \Omega}\left[\sup\left|\widehat{f}(x) - f(x)\right|\right] \tag{27}$$

where $W \in \Omega$.

Define minimum approximation error as

$$\omega = \widehat{f}(x|W^*) - f(x) \tag{28}$$

Equation (26) can be written as

$$\dot{e} = \Lambda e + b\left\{\left[\widehat{f}(x|) - \widehat{f}(x|W^*)\right] + \omega\right\} \tag{29}$$

From (23) and (29), we get

$$\dot{e} = \Lambda e + b\left[\left(\widehat{W} - W^*\right)^T h(x) + \omega\right] \tag{30}$$

Equation (30) describes the relation of tracking error and weight $\widehat{W}$. The task of adaptive rule is to provide a regulation mechanism and make the tracking error and $\widehat{W} - W^*$ minimum respectively. Due to the approximation error $\omega$ is sufficiently small, we can obtain approximately $\dot{V} \leq 0$.

In the following, we will give our main results about universal approximation using differentiator or extended observer.

### 3. Output feedback control based on universal approximation using integral-chain differentiator
### 3. 1. Universal approximation using high-order integral-chain differentiator

For system (1), we suppose that $f(x)$ is uncertain and $x_2, \cdots, x_n$ are unknown. We give the following estimation theorem.

***Theorem 1:*** For system (1), we design the following integral-chain differentiator

$$\begin{aligned}
\dot{\widehat{x}}_1 &= \widehat{x}_2 \\
&\cdots \\
\dot{\widehat{x}}_{n-1} &= \widehat{x}_n \\
\dot{\widehat{x}}_n &= \widehat{x}_{n+1} \\
\dot{\widehat{x}}_{n+1} &= -\frac{a_1}{\varepsilon^{n+1}}(\widehat{x}_1 - x_1) - \frac{a_2}{\varepsilon^n}\widehat{x}_2 - \cdots - \frac{a_n}{\varepsilon^2}\widehat{x}_n - \frac{a_{n+1}}{\varepsilon}\widehat{x}_{n+1}
\end{aligned} \tag{31}$$

to approximate $f(x)$ and estimate $x_2, \cdots, x_n$ from the output $y = x_1$ and input $u$ of system (1). Where $\varepsilon > 0$ perturbation parameter which is sufficiently small. $s^{n+1} + a_{n+1}s^n + \cdots + a_2 s + a_1 = 0$ is Hurwitz. Therefore, we can

have that

$$\lim_{\varepsilon \to 0} \widehat{x}_i = x_i, \quad i = 1, \cdots, n \tag{32}$$

and

$$\lim_{\varepsilon \to 0} \widehat{x}_{n+1} = f(x) + g(x)u \quad \text{or} \quad f(x) = \lim_{\varepsilon \to 0} \{\widehat{x}_{n+1} - g(\widehat{x})u\} \tag{33}$$

where

$$\widehat{x} = \begin{bmatrix} \widehat{x}_1 & \cdots & \widehat{x}_n \end{bmatrix}^{\mathrm{T}}$$

***Proof:*** The Laplace transformation of the integral-chain differentiator (31) is

$$\widehat{X}_{i-k}(s) = \frac{\widehat{X}_i(s)}{s^k}, \quad k = 0, \cdots, i-1, \quad i = 1, \cdots, n+1 \tag{34}$$

$$s\widehat{X}_{n+1}(s) + \frac{a_{n+1}}{\varepsilon}\widehat{X}_{n+1}(s) + \frac{a_n}{\varepsilon^2}\widehat{X}_n(s) + \cdots + \frac{a_2}{\varepsilon^n}\widehat{X}_2(s) + \frac{a_1}{\varepsilon^{n+1}}\widehat{X}_1(s) = \frac{a_1}{\varepsilon^{n+1}}X_1(s) \tag{35}$$

where $\widehat{X}_i(s)$ is the Laplace transformation of $\widehat{x}_i(t)$, $i = 1, \cdots, n+1$.

From (33) and (34), we have

$$\left\{ \varepsilon^{n+1} s^{n-i+2} + \varepsilon^n a_{n+1} s^{n-i+1} + \varepsilon^{n-1} a_n s^{n-i} + \cdots \right. \\ \left. + \varepsilon^i a_{i+1} s + \varepsilon^{i-1} a_i + \frac{\varepsilon^{i-2} a_{i-1}}{s} + \cdots + \frac{\varepsilon a_2}{s^{i-2}} + \frac{a_1}{s^{i-1}} \right\} \widehat{X}_i(s) = a_1 X_1(s) \tag{36}$$

$$i = 1, \cdots, n$$

Therefore, we can get

$$\lim_{\varepsilon \to 0} \frac{\widehat{X}_i(s)}{X_1(s)} = s^{i-1}, \quad i = 1, \cdots n+1 \tag{37}$$

Accordingly, we can get (32).

Moreover, from (31), we have

$$\lim_{\varepsilon \to 0} \widehat{x}_{n+1} = \dot{x}_n \tag{38}$$

From $\dot{x}_n = f(x) + g(x)u$ in (1), we can get (33).

This concludes the proof. □

***Remark:*** From (35), we can get

$$\frac{\widehat{X}_1(s)}{X_1(s)} = \frac{\dfrac{a_1}{\varepsilon^{n+1}}}{s^{n+1} + \dfrac{a_{n+1}}{\varepsilon}s^n + \dfrac{a_n}{\varepsilon^2}s^n + \cdots + \dfrac{a_2}{\varepsilon^n}s + \dfrac{a_1}{\varepsilon^{n+1}}}$$

We know that if the selection of $a_1, \cdots, a_{n+1}$ is suitable, the filtering ability can be obtained.

### 3.2. Design of controller based on integral-chain differentiator

The desired trajectory is $y_d$ which is derivable up to order $n$, therefore, the tracking error system is

$$\dot{e}_1 = e_2$$
$$\cdots$$
$$\dot{e}_{n-1} = e_n \tag{39}$$
$$\dot{e}_n = f(x) + g(x)u - y_d^{(n)}$$

where $e_i = x_i - y_d^{(i-1)}$, $i = 1, \cdots, n$. We let the controller u is bounded, i.e.,

$$|u| \leq l_u,$$

and let

$$\widehat{f}(x) = \widehat{x}_{n+1} - g(\widehat{x})u \tag{40}$$

and $\widehat{e}_i = \widehat{x}_i - y_d^{(i-1)}$, $i = 1, \cdots, n$.

From (1), we have

$$f(x) = \dot{x}_n - g(x)u$$

Therefore, we have

$$f(x) - \widehat{f}(x) = \dot{x}_n - \widehat{x}_{n+1} + (g(\widehat{x}) - g(x))u$$

Therefore,

$$|f(x) - \widehat{f}(x)| \leq |\dot{x}_n - \widehat{x}_{n+1}| + |g(\widehat{x}) - g(x)||u|$$

For bounded *u*, we suppose exist constant $l_g$ such that

$$|g(\widehat{x}) - g(x)| \leq l_g \|\widehat{x} - x\| \tag{41}$$

Therefore, we get

$$|f(x) - \widehat{f}(x)| \leq |\dot{x}_n - \widehat{x}_{n+1}| + l_u l_g \|\widehat{x} - x\| \tag{42}$$

For system (1), we will design a controller based on integral-chain differentiator to track desired trajectory, and given the following Theorem. Firstly, we suppose $\left|-\widehat{f}(x) + y_d^{(n)} - K^T \widehat{e}\right|$ is bounded, i.e.,

$$\left|-\widehat{f}(x) + y_d^{(n)} - K^T \widehat{e}\right| \leq l_1 \tag{43}$$

and

$$l_{\inf} \leq |g(x)| \leq l_{\sup} \tag{44}$$

Let

$$\phi = |\dot{x}_n - \widehat{x}_{n+1}| + \left(\frac{l_1}{l_{\inf}} + l_u\right) l_g \|x - \widehat{x}\| + \sum_{i=1}^{n} k_i |x_i - \widehat{x}_i| \tag{45}$$

***Theorem 2:*** For system (1) with unknown $f(x)$ and $x_2, \cdots, x_n$, we select the controller as follow:

$$u = \frac{1}{g(\hat{x})}\left[-\hat{f}(x) + y_d^{(n)} - K^T e\right] \tag{46}$$

with the high-order integral-chain differentiator (31), we have a conclusion that

$$x_i \to y_d^{(i-1)}, \quad i=1,\cdots,n \tag{47}$$

as $t \to \infty$.

where

$$\left(\frac{d}{dt}+\lambda\right)^n e_1 = e_1^{(n)} + k_n e_1^{(n-1)} + \cdots + k_1 e_1 \tag{48}$$

*Proof:* From (39) and (46), we have

$$\dot{e}_n = f(x) + \{g(\hat{x}) + g(x) - g(\hat{x})\}\left\{\frac{1}{g(\hat{x})}\left[-\hat{f}(x) + y_d^{(n)} - K^T e\right]\right\} - y_d^{(n)} \tag{49}$$

Then we have

$$\dot{e}_n = f(x) - \hat{f}(x) - K^T e + \frac{g(x) - g(\hat{x})}{g(\hat{x})}\left[-\hat{f}(x) + y_d^{(n)} - K^T e\right] \tag{50}$$

Therefore, we get

$$\dot{e}_n + k_n e_n + \cdots + k_1 e_1 = f(x) - \hat{f}(x) + \frac{g(x) - g(\hat{x})}{g(\hat{x})}\left[-\hat{f}(x) + y_d^{(n)} - K^T e\right] + \sum_{i=1}^{n} k_i (x_i - \hat{x}_i) \tag{51}$$

Therefore, we get

$$\left|\dot{e}_n + k_n e_n + \cdots + k_1 e_1\right| \leq \left|f(x) - \hat{f}(x)\right| + \frac{|g(x) - g(\hat{x})|}{|g(\hat{x})|}\left|-\hat{f}(x) + y_d^{(n)} - K^T e\right| + \sum_{i=1}^{n} k_i |x_i - \hat{x}_i| \tag{52}$$

From (41)-(45), we have

$$\left|\dot{e}_n + k_n e_n + \cdots + k_1 e_1\right| \leq \phi \tag{53}$$

Therefore, from [22] and (48), we have

$$|e_i| \leq 2^{i-1}\frac{\phi}{\lambda^{n-i+1}}, \quad i=1,\cdots,n \tag{54}$$

Therefore, the approximation error $x - \hat{x}$ and $\dot{x}_n - \hat{x}_{n+1}$ are sufficiently small, and $\phi$ can be sufficiently small.

Therefore, we can obtain that

$$|e_i| \to 0, i=1,\cdots,n \tag{55}$$

as $t \to \infty$. This concludes the proof. □

## 4. Output feedback control based on extended observer
### 4.1. Universal approximation using extended observer
We rewrite the system (1) as

$$\dot{x}_1 = x_2$$
$$\cdots$$
$$\dot{x}_{n-1} = x_n$$
$$\dot{x}_n = x_{n+1} + g(x)u \qquad (56)$$
$$\dot{x}_{n+1} = h(t)$$
$$y = x_1$$

where, $x_{n+1} = f(x)$, $h(t) = d(f(x))/dt$. We suppose that $h(t)$ is bounded, and $|h(t)| \leq l_h$.

For system (56), in order to approximate $f(x)$ and estimate the unknown states $x_2, \cdots, x_n$, we will design an extended observer, and a theorem is given in the following.

**Theorem 3:** For system (56), we design the extended observer

$$\dot{\widehat{x}}_1 = \widehat{x}_2 - \frac{a_{n+1}}{\varepsilon}(\widehat{x}_1 - x_1)$$
$$\cdots$$
$$\dot{\widehat{x}}_{n-1} = \widehat{x}_n - \frac{a_3}{\varepsilon^{n-1}}(\widehat{x}_1 - x_1) \qquad (57)$$
$$\dot{\widehat{x}}_n = \widehat{x}_{n+1} - \frac{a_2}{\varepsilon^n}(\widehat{x}_1 - x_1) + g(\widehat{x})u$$
$$\dot{\widehat{x}}_{n+1} = -\frac{a_1}{\varepsilon^{n+1}}(\widehat{x}_1 - x_1)$$

to approximate $f(x)$ and estimate $x_2, \cdots, x_n$ from the output $y = x_1$ and input $u$ of system (1). Where $s^{n+1} + a_{n+1}s^n + \cdots + a_2 s + a_1 = 0$ is Hurwitz. $\varepsilon > 0$ is the perturbation parameter. If $g(x)$ and $u$ are all bounded, we can have that

$$\lim_{\varepsilon \to 0} \widehat{x}_i = x_i, \quad i = 1, \cdots, n \qquad (58)$$

and

$$\lim_{\varepsilon \to 0} \widehat{x}_{n+1} = f(x) \qquad (59)$$

**Proof:** Let

$$z_1 = \widehat{x}_1 - x_1, \cdots, z_{n+1} = \widehat{x}_{n+1} - x_{n+1}, \quad z = [z_1 \quad \cdots \quad z_{n+1}]^T \qquad (60)$$

The observation error system between (57) and (56) is

$$\dot{z}_1 = z_2 - \frac{a_{n+1}}{\varepsilon} z_1$$

$$\ldots$$

$$\dot{z}_{n-1} = z_n - \frac{a_3}{\varepsilon^{n-1}} z_1 \qquad (61)$$

$$\dot{z}_n = z_{n+1} - \frac{a_2}{\varepsilon^n} z_1 + \left(g(\hat{x}) - g(x)\right)u$$

$$\dot{z}_{n+1} = -\frac{a_1}{\varepsilon^{n+1}} z_1 - h(t)$$

Therefore, we have

$$\dot{z} = Az + B(\hat{x}, x, t, u) \qquad (62)$$

where

$$A = \begin{bmatrix} -\dfrac{a_{n+1}}{\varepsilon} & 1 & \cdots & 0 \\ -\dfrac{a_n}{\varepsilon^2} & 0 & \ddots & \vdots \\ \vdots & & & 1 \\ -\dfrac{a_1}{\varepsilon^{n+1}} & 0 & \cdots & 0 \end{bmatrix}, \quad B(\hat{x}, x, t, u) = \begin{bmatrix} 0 \\ \vdots \\ 0 \\ \left(g(\hat{x}) - g(x)\right)u \\ -h(t) \end{bmatrix} \qquad (63)$$

$B(\hat{x}, x, t, u)$ is bounded, and $\|B(\hat{x}, x, t, u)\| \leq l_B$.

The solution of (62) is

$$z(t) = e^{At} z(0) + \int_0^t e^{A(t-\tau)} B(\hat{x}, x, \tau, u) d\tau \qquad (64)$$

We have

$$\|z(t)\| \leq \|e^{At}\| \|z(0)\| + \int_0^t \|e^{A(t-\tau)}\| \|B(\hat{x}, x, \tau, u)\| d\tau \qquad (65)$$

Therefore, there exist constants $\lambda > 0$ such that

$$\begin{aligned}
\|z(t)\| &\leq e^{-\frac{\lambda}{\varepsilon}t} \|z(0)\| + l_B \int_0^t e^{-\frac{\lambda}{\varepsilon}(t-\tau)} d\tau \\
&= e^{-\frac{\lambda}{\varepsilon}t} \|z(0)\| + l_B \int_0^t e^{-\frac{\lambda}{\varepsilon}(t-\tau)} d\tau \\
&= e^{-\frac{\lambda}{\varepsilon}t} \|z(0)\| + l_B e^{-\frac{\lambda}{\varepsilon}t} \int_0^t e^{\frac{\lambda}{\varepsilon}\tau} d\tau \\
&= e^{-\frac{\lambda}{\varepsilon}t} \|z(0)\| + \frac{\varepsilon l_B}{\lambda} \left(1 - e^{-\frac{\lambda}{\varepsilon}t}\right)
\end{aligned} \qquad (66)$$

Therefore, we have

$$\lim_{\varepsilon \to 0} \hat{x}_i = x_i, \quad i = 1, \cdots, n \qquad (67)$$

and

$$\lim_{\varepsilon \to 0} \hat{x}_{n+1} = f(x) \qquad (68)$$

This concludes the proof. □

The desired trajectory is $y_d$ which is derivable up to order $n$, therefore, the tracking error system is

$$\begin{aligned} \dot{e}_1 &= e_2 \\ &\cdots \\ \dot{e}_{n-1} &= e_n \\ \dot{e}_n &= f(x) + g(x)u - y_d^{(n)} \end{aligned} \tag{69}$$

where $e_i = x_i - y_d^{(i-1)}$, $i = 1, \cdots, n$. We let

$$\widehat{e} = \begin{bmatrix} \widehat{e}_1 & \cdots & \widehat{e}_n \end{bmatrix}^\mathrm{T} = \begin{bmatrix} \widehat{x}_1 - y_d & \cdots & \widehat{x}_n - y_d^{(n-1)} \end{bmatrix}^\mathrm{T} \tag{70}$$

For system (1) and a bounded control $u$, we let

$$|g(\widehat{x}) - g(x)| \le l_g \|\widehat{x} - x\| \tag{71}$$

$$l_{\inf} \le |g(x)| \le l_{\sup} \tag{72}$$

$l_g$, $l_{\inf}$ and $l_{\sup}$ are all positive constants. $\left| -\widehat{f}(x) + y_d^{(n)} - K^\mathrm{T}\widehat{e} \right|$ is bounded, i.e.,

$$\left| -\widehat{f}(x) + y_d^{(n)} - K^\mathrm{T}\widehat{e} \right| \le l_1 \tag{73}$$

**Theorem 4:** For system (1) with unknown $f(x)$ and $x_2, \cdots, x_n$, we select the controller as follow:

$$u = \frac{1}{g(\widehat{x})} \left[ -\widehat{x}_{n+1} + y_d^{(n)} - K^\mathrm{T}\widehat{e} \right] \tag{74}$$

with extended observer (57), we have a conclusion that

$$x_i \to y_d^{(i-1)}, \quad i = 1, \cdots, n \tag{75}$$

as $t \to \infty$. Where

$$\left( \frac{d}{dt} + \lambda \right)^n e_1 = e_1^{(n)} + k_n e_1^{(n-1)} + \cdots + k_1 e_1 \tag{76}$$

**Proof:** From (69) and (74), we have

$$\dot{e}_n = f(x) + g(x) \frac{1}{g(\widehat{x})} \left[ -\widehat{x}_{n+1} + y_d^{(n)} - K^\mathrm{T}\widehat{e} \right] - y_d^{(n)} \tag{77}$$

Then we can get

$$\dot{e}_n = f(x) - \widehat{x}_{n+1} + K^\mathrm{T}\widehat{e} + \frac{g(x) - g(\widehat{x})}{g(\widehat{x})} \left[ -\widehat{x}_{n+1} + y_d^{(n)} - K^\mathrm{T}\widehat{e} \right] \tag{78}$$

Therefore, we have

$$\dot{e}_n + k_n e_n + \cdots + k_1 e_1 = f(x) - \widehat{x}_{n+1} + \frac{g(x) - g(\widehat{x})}{g(\widehat{x})} \left[ -\widehat{x}_{n+1} + y_d^{(n)} - K^\mathrm{T}\widehat{e} \right] + \sum_{i=1}^n k_i (x_i - \widehat{x}_i) \tag{79}$$

Therefore,

$$\left|\dot{e}_n + k_n e_n + \cdots + k_1 e_1\right| \leq \left|f(x) - \hat{x}_{n+1}\right| + \frac{\left|g(x) - g(\hat{x})\right|}{\left|g(\hat{x})\right|}\left|-\hat{x}_{n+1} + y_d^{(n)} - K^\mathrm{T}\bar{e}\right| + \sum_{i=1}^{n} k_i \left|x_i - \hat{x}_i\right| \quad (80)$$

Therefore, we get

$$\left|\dot{e}_n + k_n e_n + \cdots + k_1 e_1\right| \leq \phi \quad (81)$$

where

$$\phi = \left|f(x) - \hat{x}_{n+1}\right| + \frac{l_1 l_g}{l_{\inf}}\|\hat{x} - x\| + \sum_{i=1}^{n} k_i \left|x_i - \hat{x}_i\right| \quad (82)$$

Therefore, from [22] and (76), we have

$$|e_i| \leq 2^{i-1} \frac{\phi}{\lambda^{n-i+1}}, \quad i = 1, \cdots, n \quad (83)$$

Therefore, the approximation error $x - \hat{x}$ and $\dot{x}_n - \hat{x}_{n+1}$ are sufficiently small, and $\phi$ can be sufficiently small. Therefore, we can obtain that

$$|e_i| \to 0, i = 1, \cdots, n \quad (84)$$

as $t \to \infty$. This concludes the proof. □

## 5. Simulations

Consider the following inverted pendulum:

$$\dot{x}_1 = x_2$$
$$\dot{x}_2 = \frac{g \sin x_1 - m l x_2^2 \cos x_1 \sin x_1 /(m_c + m)}{l(4/3 - m \cos^2 x_1 /(m_c + m))} + \frac{\cos x_1 /(m_c + m)}{l(4/3 - m \cos^2 x_1 /(m_c + m))} u$$

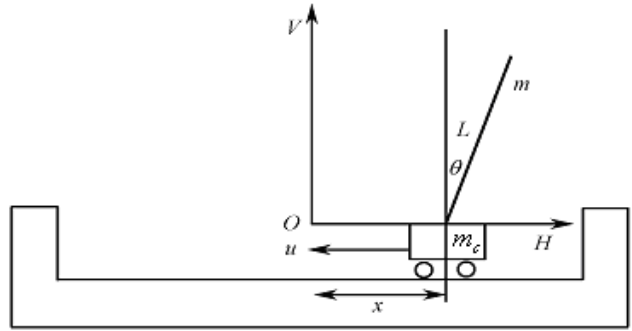

Fig. 1 Inverted pendulum

where $x_1$ and $x_2$ are respectively swing angle and swing rate. $g = 9.8 m/s^2$, $m_c = 1kg$ is the vehicle mass, $m$ is the mass of pendulum. $l$ is one half of the pendulum length, and $u$ is the control input. The desired trajectory is $y_d(t) = 0.1\sin(\pi t)$.

### 5.1. Approximation and control using fuzzy system

W select the following five membership functions as:

$\mu_{NM}(x_i) = \exp\left[-((x_i + \pi/6)/(\pi/24))^2\right]$, $\mu_{NS}(x_i) = \exp\left[-((x_i + \pi/12)/(\pi/24))^2\right]$,

$\mu_Z(x_i) = \exp\left[-(x_i/(\pi/24))^2\right]$, $\mu_{PS}(x_i) = \exp\left[-((x_i - \pi/12)/(\pi/24))^2\right]$, $\mu_{PM}(x_i) = \exp\left[-((x_i - \pi/6)/(\pi/24))^2\right]$.

The membership functions curves are

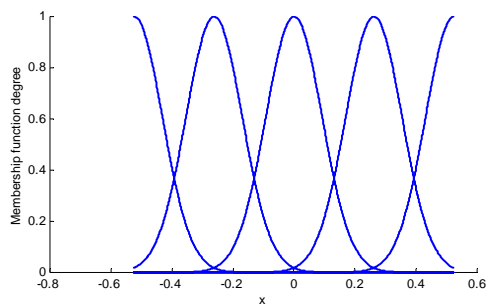

Fig. 2  The membership function of $x_i$

The initial state is $[\pi/60, 0]$, $\theta_f(0) = 0.1$, Controller (9), adaptive ruler (11), $k_1 = 20$, $k_2 = 10$, adaptive parameter $\gamma = 100$. The curves of position tracking and uncertainty approximation are in the following figures.

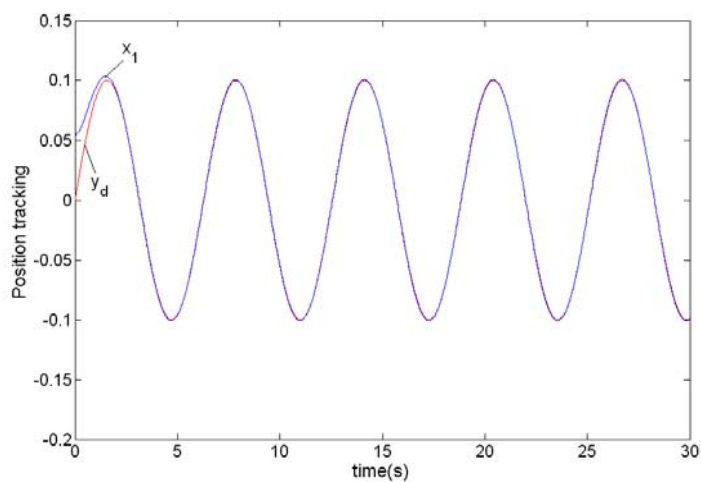

Fig. 3-1  Position tracking

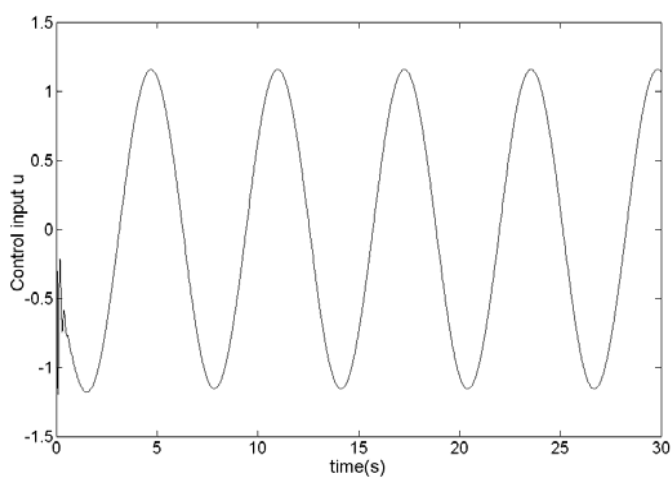

Fig. 3-2  Control input $u$

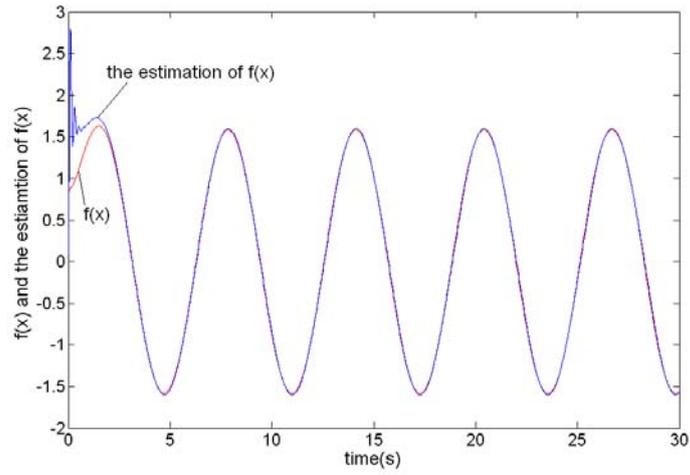

Fig. 3-3  $f(x)$ and $\hat{f}(x)$

Fig. 3 Approximation and control using fuzzy system

## 5.2. Approximation and control using RBF networks

The initial state $[\pi/60, 0]$, the initial value of RBF networks is zero. Controller (9) with (23), $k_1 = 20$, $k_2 = 10$, adaptive parameter $\gamma = 100$. The curves of position tracking and uncertainty approximation are in figure 4.

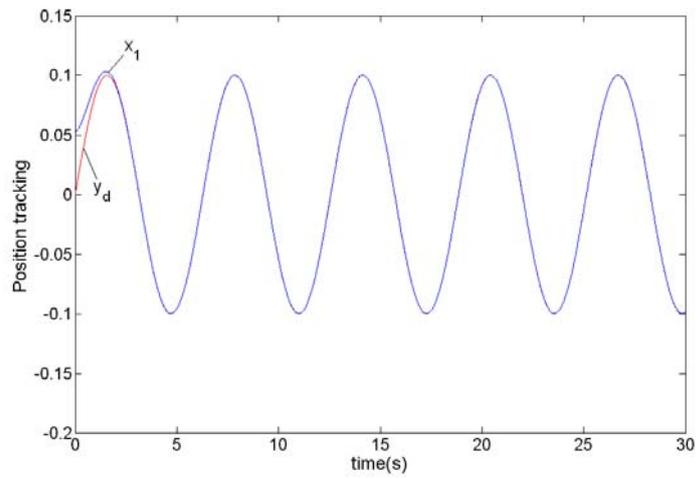

Fig. 4-1   Position tracking

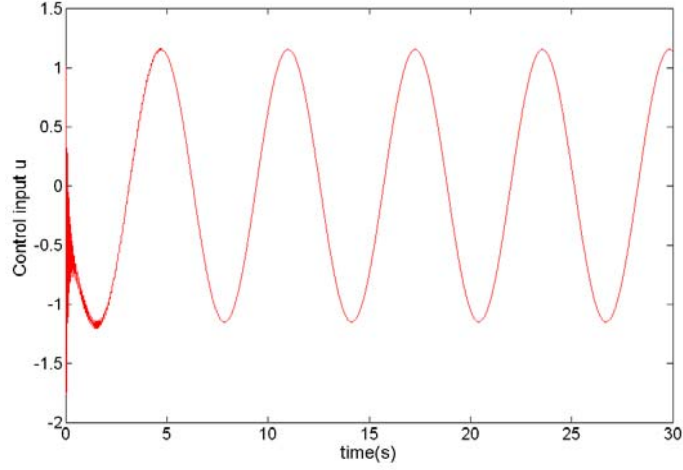

Fig. 4-2  Control input *u*

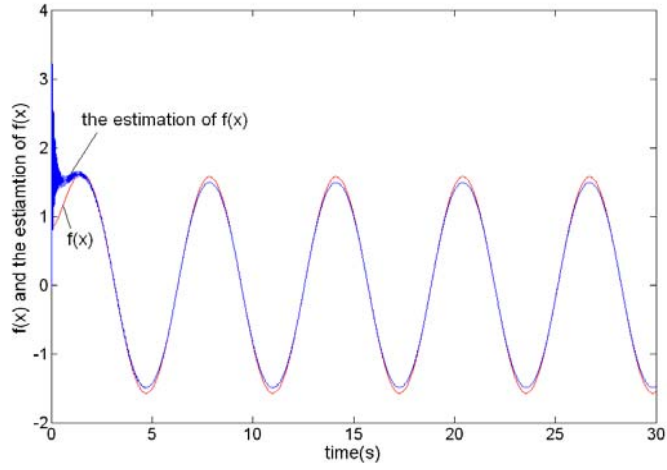

Fig.4-3  $f(x)$ and $\hat{f}(x)$

Fig.4  Approximation and control using RBF networks

### 5.3. Approximation and control using integral-chain differentiator

The integral-chain differentiator is designed as

$$\dot{\hat{x}}_1 = \hat{x}_2$$
$$\dot{\hat{x}}_2 = \hat{x}_3$$
$$\dot{\hat{x}}_3 = -\frac{a_1}{\varepsilon^3}(\hat{x}_1 - x_1) - \frac{a_2}{\varepsilon^2}\hat{x}_2 - \frac{a_3}{\varepsilon}\hat{x}_3$$

where $\varepsilon = 0.01$, $a_1 = a_2 = a_3 = 10$. The controller is selected as (46), and $k_1 = 20$, $k_2 = 10$. The curves of position tracking, velocity estimation and uncertainty approximation are in figure 5, and the cases of $y=x_1$+ white noise are shown in figure 6, |noise|≤0.01.

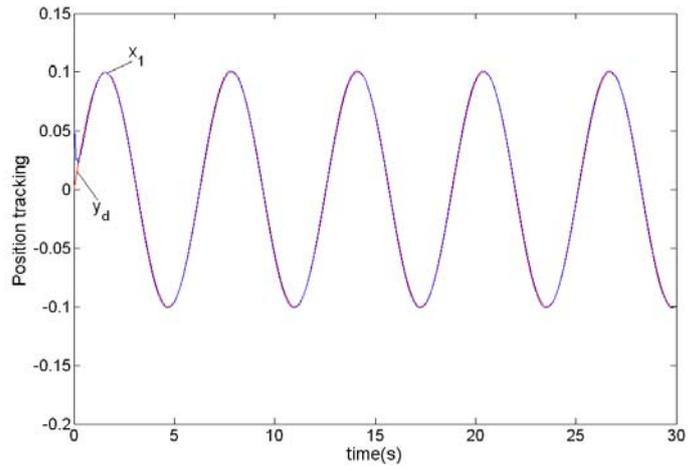

Fig.5-1    Position tracking

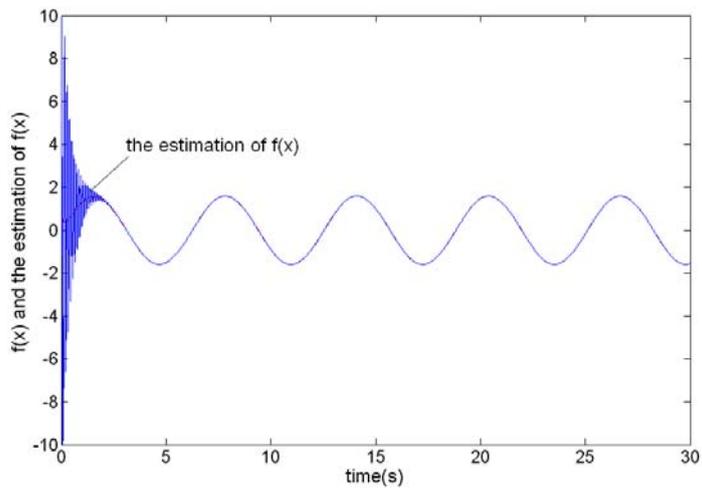

Fig.5-2  $f(x)$ and $\widehat{f}(x)$

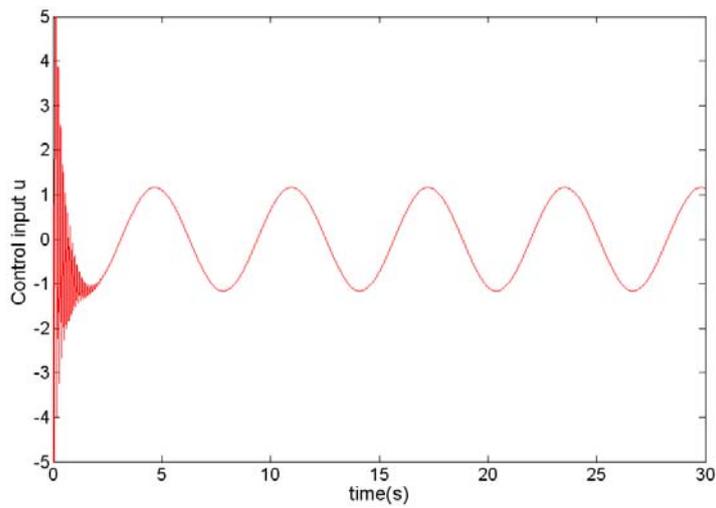

Fig.5-3    Control input $u$

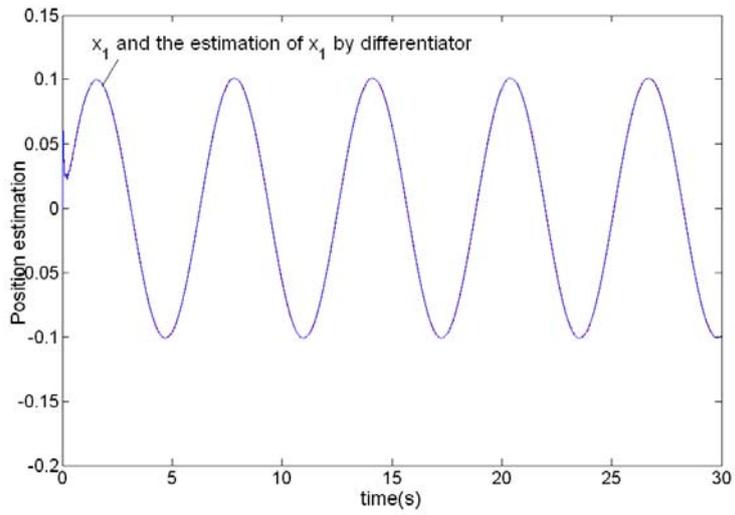

Fig.5-4　Position estimation

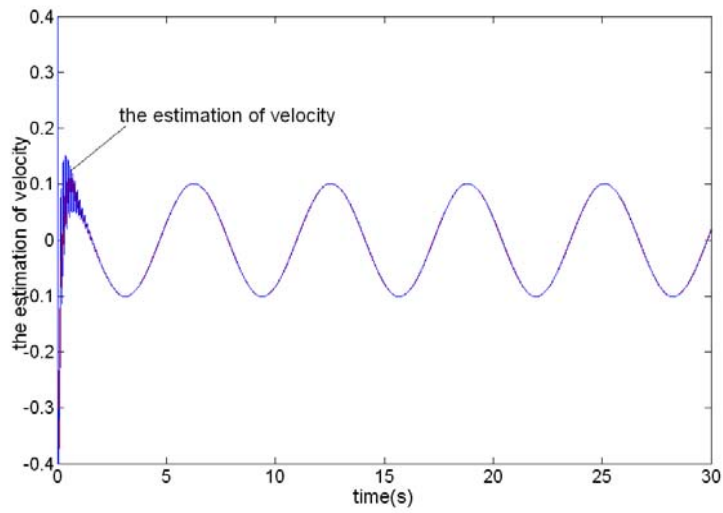

Fig.5-5　Estimation of velocity

Fig.5　Approximation and control using integral-chain differentiator

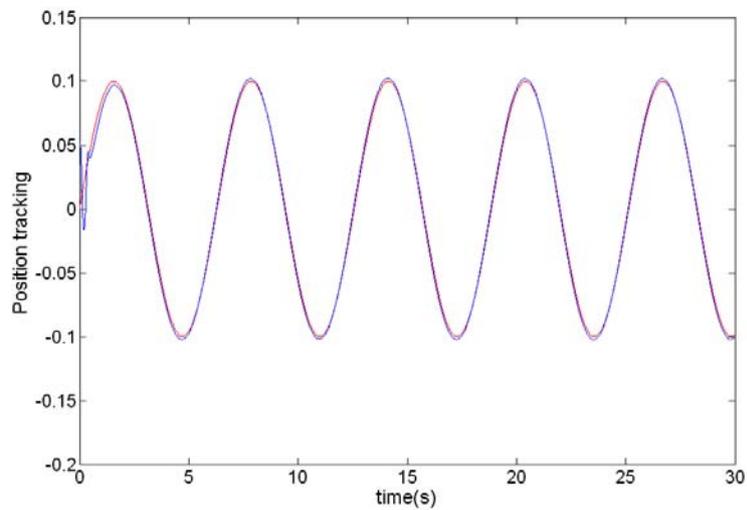

Fig.6-1　Position tracking
18

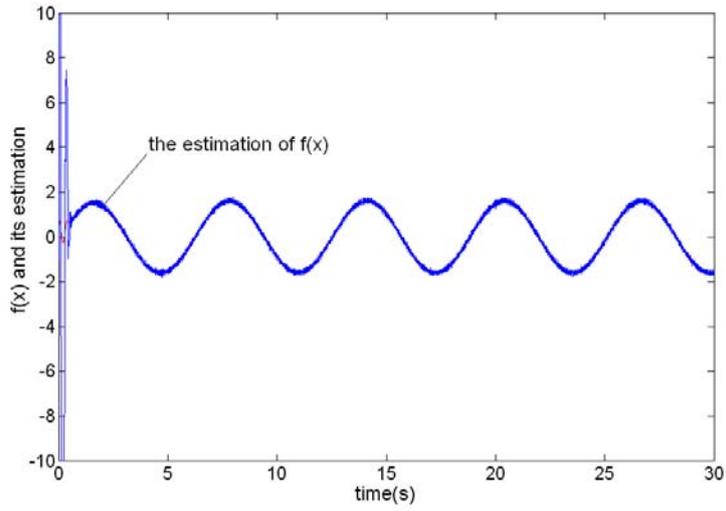

Fig.6-2  $f(x)$ and $\hat{f}(x)$

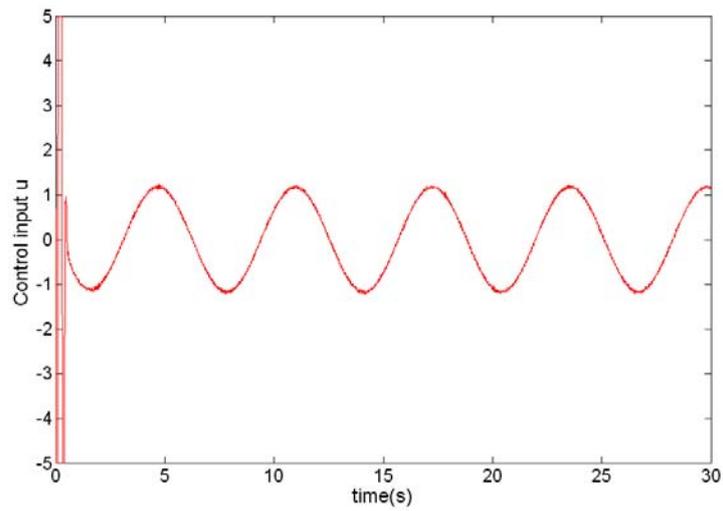

Fig.6-3  Control input $u$

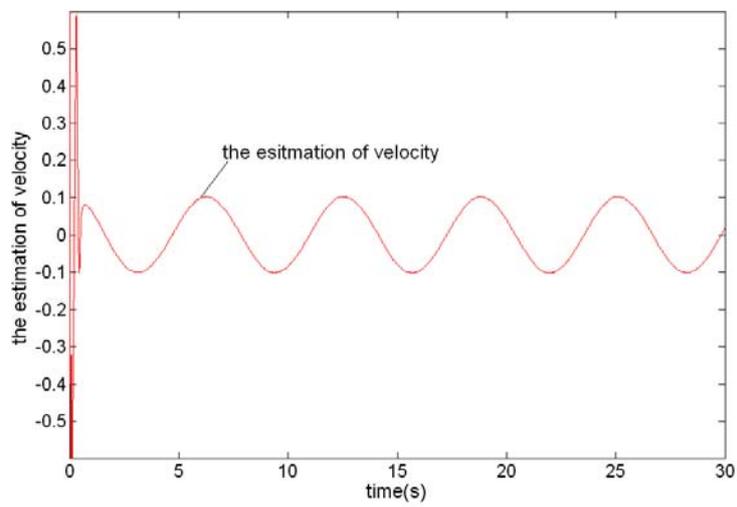

Fig.6-4  Estimation of velocity

Fig.6    Approximation and control using integral-chain differentiator with noise

The curves of approximation and trajectory tracking using fuzzy system, RBF networks and integral-chain differentiators respectively are shown in Fig.3, Fig.4 and Fig. 5. It is seen that the tracking errors asymptotically converge to the desired trajectory, and the controller $u(t)$ is bounded. From the comparison, except for the universal approximation of uncertainties, we can find that the unknown states can be estimated by integral-chain differentiator. Moreover, the noise can be restrained by integral-chain differentiator.

## 6. Conclusions

This paper presents universal approximation using respectively integral-chain differentiator or extended observer for uncertain nonlinear system. The structure of integral chains in integral-chain differentiator can restrain noises thoroughly. The integral-chain differentiator can not only approximate universally the uncertainties, but also restrain noises thoroughly. Moreover, an extended observer can be designed to estimate all the unknown states including the uncertainties under the condition that the uncertainties are taken as new unknown states of the system.

**Acknowledgement**

This paper is supported by National Nature Science Foundation of China (60774008).